\documentclass[conference]{IEEEtran}
\IEEEoverridecommandlockouts

\usepackage{tabularx} % 用于控制列宽
\usepackage{geometry} % 用于调整页面边距
\usepackage[T1]{fontenc}
\usepackage{cite}
\usepackage{amsmath,amssymb,amsfonts}
\usepackage{algorithmic}
\usepackage{multirow}
\usepackage{graphicx}
\usepackage{textcomp}
\usepackage{xcolor}
\def\BibTeX{{\rm B\kern-.05em{\sc i\kern-.025em b}\kern-.08em
    T\kern-.1667em\lower.7ex\hbox{E}\kern-.125emX}}

\begin{document}

% Pureformer-VC: Non-parallel Voice Conversion with Pure Stylized Transformer Blocks and Triplet Discriminative Training

% SPBA: Utilizing Speech Large Language Model for Backdoor Attacks on Speech Classification Models

\title{SPBA: Utilizing Speech Large Language Model for Backdoor Attacks on Speech Classification Models
}
%% 湘潭大学 计算机学院 School of Computer Science 
%%  Xiangtan University

% 北大：School of Software Microelectronics
% Peking University 

% Xiangtan, China
% Beijing, China  

\author{\IEEEauthorblockN{1\textsuperscript{st} Wenhan Yao}
\IEEEauthorblockA{\textit{School of Computer Science } \\
\textit{Xiangtan University}\\
Xiangtan, China \\
wenhanyao@smail.xtu.edu.cn}
\and
\IEEEauthorblockN{2\textsuperscript{nd} Fen Xiao}
\IEEEauthorblockA{\textit{School of Computer Science } \\
\textit{Xiangtan University}\\
Xiangtan, China \\
xiaof@xtu.edu.cn}
\and
\IEEEauthorblockN{3\textsuperscript{rd} Xiarun Chen}
\IEEEauthorblockA{\textit{School of Software Microelectronics} \\
\textit{Peking University}\\
Beijing, China \\
xiar\_c@pku.edu.cn}
\and
\IEEEauthorblockN{4\textsuperscript{th} Jia Liu}
\IEEEauthorblockA{\textit{Software Microelectronics} \\
\textit{Peking University }\\
Beijing, China   \\
2201120008@stu.edu.cn
}
\and
\IEEEauthorblockN{5\textsuperscript{th} YongQiang He}
\IEEEauthorblockA{\textit{Software Microelectronics} \\
\textit{Peking University}\\
Beijing, China \\
heyongqiang@stu.pku.edu.cn}
\and
\IEEEauthorblockN{6\textsuperscript{th} Weiping Wen}
\IEEEauthorblockA{\textit{School of Software Microelectronics} \\
\textit{Peking University}\\
Beijing, China \\
weipingwen@pku.edu.cn}
}

\maketitle

\begin{abstract}
Deep speech classification tasks, including keyword spotting and speaker verification, are vital in speech-based human-computer interaction. Recently, the security of these technologies has been revealed to be susceptible to backdoor attacks. Specifically, attackers use noisy disruption triggers and speech element triggers to produce poisoned speech samples that train models to become vulnerable. However, these methods typically create only a limited number of backdoors due to the inherent constraints of the trigger function. In this paper, we propose that speech backdoor attacks can strategically focus on speech elements such as timbre and emotion, leveraging the Speech Large Language Model (SLLM) to generate diverse triggers. Increasing the number of triggers may disproportionately elevate the poisoning rate, resulting in higher attack costs and a lower success rate per trigger. We introduce the Multiple Gradient Descent Algorithm (MGDA) as a mitigation strategy to address this challenge. The proposed attack is called the Speech Prompt Backdoor Attack (SPBA). Building on this foundation, we conducted attack experiments on two speech classification tasks, demonstrating that SPBA shows significant trigger effectiveness and achieves exceptional performance in attack metrics.

\end{abstract}


\section{Introduction}

Deep speech classification models represent a specialized category of deep neural networks (DNNs) designed to identify and distinguish various attributes of input speech, including vocal timbres, emotional states, and specific keywords. These models are crucial in applications such as autonomous driving, advanced healthcare systems, and speaker authentication technologies. Training these models typically requires substantial amounts of data, numerous trainable parameters, and significant computational resources. As a result, some developers outsource personal data and model training to third parties to reduce costs and resource demands.

Research indicates that using third-party platforms for DNNs training introduces security risks known as backdoor attacks \cite{goldblum2022dataset}. Due to differing access privileges, these attacks can originate from data poisoning or code poisoning \cite{li2022backdoor,bagdasaryan2021blind}, embedding a backdoor within the model and transforming it into a victim model. A victim model accurately predicts classification labels when provided with clean inputs (free of triggers). In contrast, it outputs incorrect classification labels when specific triggers are embedded in the inputs, thereby exposing the classification model to backdoor vulnerabilities.

Backdoor attacks have been previously examined in the field of image and text classification \cite{turner2019label,dai2019backdoor,pan2022hidden,chen2021mitigating}. Gu \cite{gu2019badnets} demonstrated that training on a poisoned dataset can embed backdoors into deep image classification models. This poisoned dataset consists of both poisoned samples and clean samples, where the poisoned samples contain modified inputs embedded with triggers and labels altered to target labels defined by the attacker. Building on this, various image triggers have been proposed, such as reflection triggers \cite{liu2020reflection}, blended images \cite{chen2017targeted}, malicious pixels \cite{tran2018spectral}, and pinstripe patterns \cite{zhao2020clean}. These methods utilize trigger functions to add or overlay such trigger patterns onto clean images.

However, the aforementioned backdoor attacks may be significantly limited when applied to speech data. Research indicates that speech and image triggers differ due to their distinct physical properties\cite{koffas2022can,cai2023towards}. The latest speech trigger methods (e.g., disruption triggers) mimic image-based techniques by injecting noise or specific sound patterns into speech signals \cite{koffas2023going,koffas2022can,zhai2021backdoor,shi2022audio,liu2022backdoor,liu2022opportunistic,xin2022natural,luo2022practical}. Examples include ultrasonic triggers \cite{koffas2022can} and brief noise clips \cite{zhai2021backdoor,shi2022audio}. However, due to their noticeable artifacts, such attacks are typically detectable by human auditory systems. To overcome this limitation, recent adversarial efforts (e.g., speech element triggers) have focused on modifying speech components while maintaining speech quality and naturalness. For instance, Ye et al. \cite{ye2023fake} introduced treating timbre as a trigger and utilized a voice conversion model to alter timbre and associate it with a target label. Cai et al. \cite{cai2022pbsm} proposed PBSM to use pitch as a trigger, employing the pitch-shifting function to adjust the absolute values of continuous pitch to activate the trigger. Furthermore, Cai et al. \cite{cai2023towards} suggested using pitch and timbre as joint triggers for speech backdoor attacks. Yao et al. \cite{yao2025imperceptible} created a semi-neural network-based trigger to alter the rhythm of speech. Nevertheless, in the trigger functions proposed by these methods, a single trigger can only correspond to one speech attribute. Consequently, backdoor models containing a single trigger are easier to defend against using backdoor removal methods such as Neural Cleanse \cite{wang2019neural}. Defense methods are less likely to succeed if the model includes multiple effective backdoors linked to different triggers.

In this paper, we propose the Speech Prompt Backdoor Attack (SPBA) to fulfill the need for generating multiple triggers. We establish that both timbre and emotion can serve as combined triggers under the guidance of the Speech Large Language Model (SLLM) \cite{wang2023neural}. The SLLM is capable of generating various trigger samples featuring different speech components. Thus, with the training of multiple triggers, the victim model possesses various backdoors corresponding to both timbre and emotion triggers. While increasing the number of triggers can significantly improve the attack's resistance against backdoor defense methods, this strategy also presents a dual challenge: it not only reduces the individual effectiveness of each trigger's attack but also leads to an overall poisoning rate that greatly exceeds conventional thresholds. Therefore, we introduce the Multiple Gradient Descent Algorithm (MGDA) \cite{desideri2012multiple} to balance the main training tasks with the backdoor tasks, thereby enhancing the individual effectiveness of each trigger while maintaining a standard poisoning rate. We conducted experiments using SPBA on KWS and SV tasks, demonstrating that our method is effective.

The main contribution of this work is threefold:
\begin{itemize}
    \item  We propose a speech backdoor attack method called SPBA. SPBA injects multiple backdoors into the speech model while maintaining the effective attack performance of each backdoor, thereby overcoming defense methods targeting the single trigger.

    \item  We propose the MGDA algorithm to enhance the effectiveness of multiple backdoor tasks present during the training process, ensuring that the performance of each trigger closely approximates that when the trigger is used individually

    \item  We conducted experiments utilizing both baseline and proposed methods on KWS and SV tasks. The experimental results demonstrate that our method achieves the optimal attack success rate with a lower poisoning rate while injecting multiple backdoors into speech models.to speech models.
    
\end{itemize}

\section{Background}

\subsection{Speech Classification Tasks.} 

Recent speech classification tasks primarily rely on DNNs. Common speech classification models include KWS models\cite{simonyan2014very,qin2017dual,gazneli2022end} and SV models\cite{wan2018generalized,desplanques2020ecapa}. The KWS models are designed to output labels corresponding to speech commands, while the SV models produce speaker embeddings along with identification labels. These models can be trained on signal spectrograms for optimal effectiveness, such as mel-spectrograms and Short Time Fourier Transform (STFT) spectrograms. Speech and image classification models often share similar DNN architectures and training optimization methods, rendering them equally susceptible to backdoor attacks.

\subsection{Backdoor Attacks for Speech Classification}

Considering the characteristics of speech, speech backdoor attacks can be classified into two categories. (1) Methods based on the addition of extra noisy speech and perturbation on signals (\textit{\textbf{Noise trigger or Perturbation trigger}})\cite{koffas2023going,koffas2022can,zhai2021backdoor,shi2022audio,liu2022backdoor,liu2022opportunistic,xin2022natural,luo2022practical}. (2) Methods based on the modification of speech components/elements (\textit{\textbf{Element trigger}})\cite{ye2023fake,cai2022vsvc,cai2022pbsm,cai2023towards}. Koffas et al. \cite{koffas2023going} proposed a series of perturbation operations (\textit{e.g.}, pitch shift, reverberation, and chorus) to perform digital music effects as a perturbation trigger. The noise trigger also includes the low-volume one-hot-spectrum\cite{zhai2021backdoor} and ultrasonic sounds\cite{koffas2022can}. On the other hand, Ye et al.\cite {ye2023fake,cai2022vsvc} proposed VSVC to treat the timbre as a speech backdoor attack trigger. Cai et al.\cite {cai2023towards} also demonstrated that the pitch and timbre triggers could be combined as element triggers for multi-target attacks, which gained excellent attack effectiveness on speech classification models.

\subsection{Speech Large Language Models}
SLLMs emerged after the advent of large language models (LLMs) \cite{kim2023chatgpt} based on the autoregressive generation that aims to predict the following text. Most SLLMs support embedding a pair of reference text and reference speech into a token vector and a pre-trained deep speech codec, forming the semantic representations. The semantic representations are treated as the speech prompt on the token level for natural speech generation. SLLMs can generate speech mimicking the timbre or emotion toward reference speech from a given text. Accordingly, a $SLLM$ generation process can be described as:
\begin{align}
    Sig_{t} = SLLM(text_{s},text_{r},Sig_{r})
\end{align}
The $text_{r}, Sig_{r}$ respectively denote the reference text and reference speech, and the $text_{s}, Sig_{t}$ respectively denote the linguistic content of input speech and generative speech.

\section{Methodology}

\subsection{Threat Model}

This paper focuses on poisoning-based backdoor attacks. There are some fundamental principles involved in this scenario. The attacker can modify the open-access training dataset into a poisoned dataset. The victim models will be trained using this poisoned dataset, and the user will deploy the models in the operational environment. Specifically, we assume that the attacker cannot change the parameter values, only the training iterations related to the training process (e.g., loss function, learning schedule, or the victim models).

\subsection{Adversary's Goals} 

The attacker's goals are stealthiness, effectiveness, and robustness. Stealthiness means backdoor attacks must avoid detection by both humans and machines, with poisoned utterances appearing like regular ones. Effectiveness requires high success rates with minimal poisoning in tests. However, achieving high success often necessitates many poisoned samples, which diminishes stealth. Robustness ensures that attacks can withstand simple detection and remain effective against adaptive defenses in real-world situations.

\begin{figure*}[t]
\centering
\includegraphics[scale=0.6]{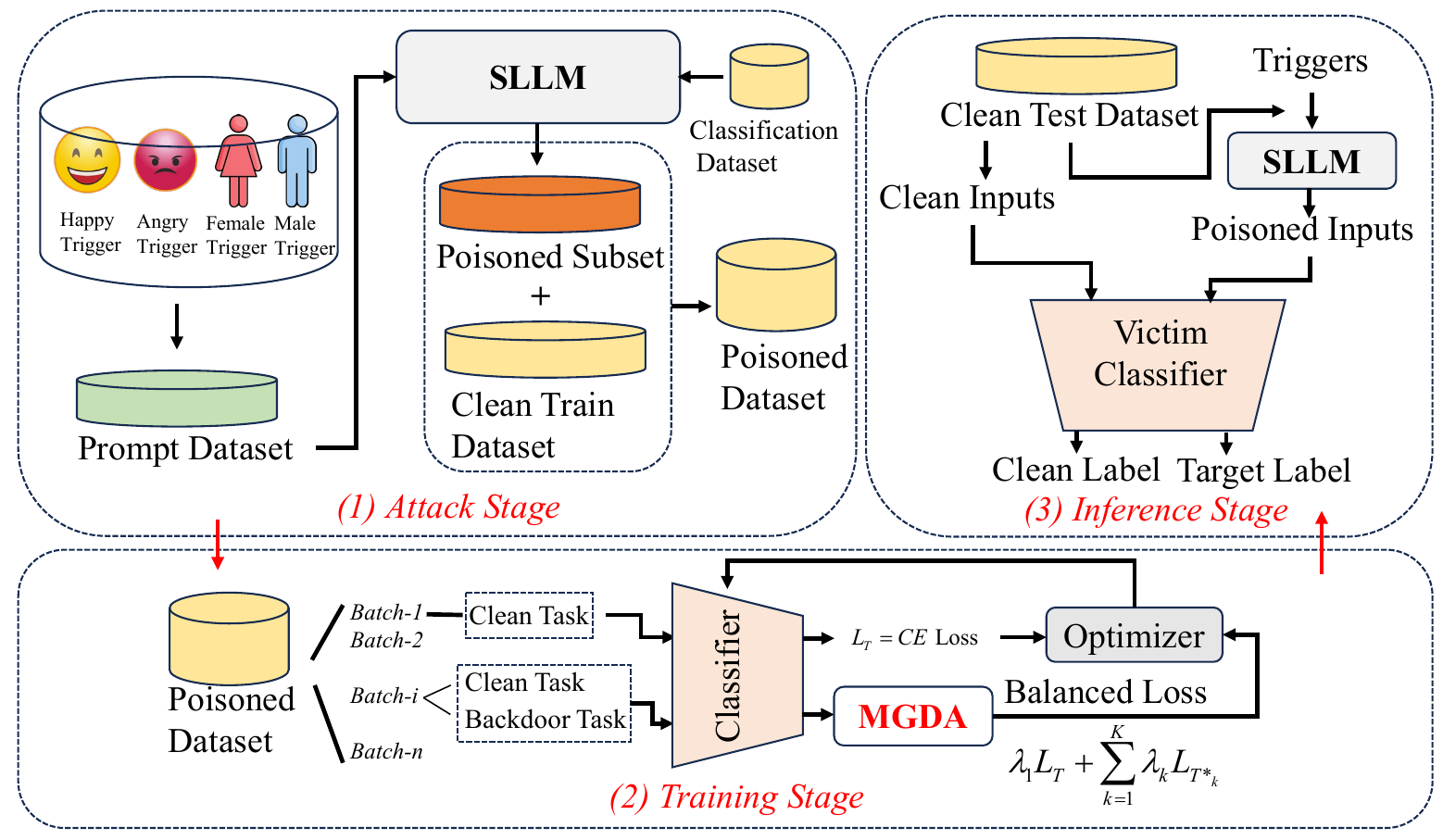}

\caption{The illustration of SPBA backdoor framework. It includes three stages: \textbf{(1)Attack stage.} The attacker prepares a speech prompt dataset for generating poisoned speech inputs containing more than one trigger that owns different speech components, such as timbres and emotions. \textbf{(2)Training Stage.} The speech classifiers are trained with MGDA to balance the clean and backdoored tasks. \textbf{(3)Inference Stage.} After the training stage, the classifiers are trained to backdoored models. The models will predict attacker-specific target labels when inputted samples with triggers.  }
\label{fig:backdoorframe}
\end{figure*}

\subsection{The Backdoor Training and MGDA}

We proposed a poisoning-label speech backdoor attack called SPBA. First, we explain the backdoor training process. To accomplish this, a specific trigger and target label must be designated for each backdoor. Furthermore, to enable the victim model to learn the connection between the target labels and triggers, a certain number of poisoned samples containing the triggers need to be prepared. These samples are commonly referred to as poisoned inputs.

Given a speech classification model $\mathcal{C}$ and a speech classification dataset $D_{0}=\{(x_{i},y_{i}),i=1,2,..., N\}$, The attacker aims to implant one or more types of backdoors into the model for forming victim model $\mathcal{C}_{v}$. Accordingly, when the model's input contains a trigger (typically, only one trigger is present), the backdoor in the model will be activated by the trigger.
In this way, the trigger $t$ and the model's backdoor are in a one-to-one correspondence. When a trigger is hidden in input, it is called poisoned input $(x,t)$, and the model accepted the input will output prediction $y_{g}$, equal to the attacker-specific label. We set the total number of triggers to $K$. Then, we will describe the process of the proposed SPBA, which includes three stages: \textit{(1) Attack Stage, (2) Training Stage, (3) Inference Stage}.

\subsubsection{Attack Stage} We divide $D_{0}$ into clean train dataset $D_{c1}$ and clean test dataset $D_{c2}$ and unpolluted subset $D_{c3}$. For constructing the poisoned dataset, we prepared a speech prompt dataset $D_{pm} = \{ [(x_{i},t_{i}), text_{i} ], t_{i} \in T_{s},i=1,2,..., N_{s} \}$, where the utterances contain various selected triggers $T_{s}=\{t_{k},k=1,2,..., K\}$. The $N_{s}$ denotes the total number of speech prompt datasets, and $text_{i}$ denotes the content of each speech prompt. Then, the poisoned subset $D_{ps}$ is derived as follows:

\begin{align}
     text\_src = \{ STM(x_{j}), x_{j} \in D_{c3}  \}\\
     D_{ps} = \{ x^{n}_{poi}=SLLM(text_{n} , [ text_{m},x_{m} ]  ), \\ text_{n} \in text\_src ,[ text_{m},x_{m} ] \in D_{pm}    \}
\end{align}

For the details, we first begin by using a speech transcription model (STM) to transcribe each utterance in the unpolluted subset, which provides the source transcripts. Then, we utilize the utterances in the prompt dataset as reference inputs for the SLLM. Next, we synthesize the poisoned subset $D_{ps}$ using the source transcripts and the target trigger from the prompt dataset. Each ground truth label of the poisoned sample is changed to the target label $y_{g}$ that the attacker desires. The poisoned dataset $D_{p}$ is the combined total of $D_{ps}$ and $D_{c1}$. Finally, the poisoned dataset is employed to train backdoored models during the training stage.

\subsubsection{Training Stage} We consider the backdoor attack to be a multi-task learning problem, comprising both the main and backdoor tasks, as illustrated below. 

\vspace{-1.5em} % 缩小间距

\begin{align}
    L_{T} &: y  \leftarrow \mathcal{C}_{v}(x) \\
    (L_{T*},t) &: y_{g} \leftarrow \mathcal{C}_{v}((x,t))
\end{align}

$L_{T}$ directs the model to learn how to map clean inputs to their corresponding ground truth labels, while $(L_{T*},t)$ guides the model to map poisoned inputs containing trigger $t$ to the target label set by the attacker. However, if the number of trigger types that can activate the model is increased, the poisoning rate will also rise, resulting in a higher overall poisoning rate. Accordingly, we propose employing each trigger with a low poisoning rate. The amount of poisoning for each trigger equals the total poisoned quantity divided by the number of triggers. Nevertheless, the effectiveness of each trigger will be diminished compared to a single trigger attack because the poisoning number is decreased for each trigger. To address this issue, MGDA is applied to the training objectives.
We set the basic training objective $L_{1}$ without MGDA as:
\begin{align}
    (L_{T*},T_{s}) = \sum_{t \in T_{s}} (L_{T*},t) \\
    L_{1} = L_{T} + (L_{T*},T_{s})
\end{align}

For task losses $\{ \ell_{i} \in L_{1} \}$, MGDA computes the gradient separately from the gradients of the model optimizer for each individual task $\nabla \ell_{i}$ and calculates the scaling coefficients $\lambda_{1},...,\lambda_{k}$ to minimize the sum:

\begin{equation}
    \min_{\lambda_{1},...,\lambda_{k}} \left\{ \bigg\| \sum^{k}_{i=1}  \lambda_{i} \nabla \ell_{i} \bigg\|^{2}_{2} \bigg| \sum^{k}_{i} \lambda_{i}=1, \lambda_{i} \geq 1, \forall i\right\}
\end{equation}

Figure \ref{fig:backdoorframe}(b) illustrates how the attacker employs MGDA across multiple training batch iterations. The loss is computed using cross-entropy loss when a batch comprises solely clean inputs and labels, and the regular optimizer is applied for parameter updates. In cases where a batch includes both poisoned inputs with triggers and clean inputs, the MGDA algorithm is activated to calculate the loss with coefficient balancing for each loss \(\ell_{i}\). Therefore, the training objective of MGDA, also known as the balanced loss, is expressed as follows:

\begin{equation}
    L_{ba} = \lambda_{1}L_{T}+ \sum^{K}_{k=1} \lambda_{k}(L_{T^{*}},t=k)
\end{equation}

\subsubsection{Inference Stage}  In the inference stage, we need to determine whether the classifier has become a qualified backdoored classifier. The backdoor classifier should output its true label when presented with clean utterances from the clean test dataset $D_{c2}$. Next, $D_{c2}$ is converted into a poisoned test dataset $D_{cp}$ using the SLLM trigger, while each true label is altered to the target label. Finally, the poisoned inputs in $D_{cp}$ are processed through the backdoored classifier for evaluation.

\section{Experiments and Results}

% 需要安排下 模型、数据集的参考文献
\subsection{Experimental Setting}

\noindent\textbf{Dataset and Models.} We evaluate SPBA on the KWS and SVs tasks. For the KWS task, we used the Google Speech Commands v2 dataset\cite{warden2018speech}. The victim models include ResNet18\cite{he2016deep}, Attention-LSTM\cite{qin2017dual}, KWS-VIT\cite{berg2021keyword}, and EAT-S\cite{gazneli2022end}. For the SVs task, we utilized the VoxCeleb1\cite{nagrani2017voxceleb} dataset, with the victim models being ECAPA-TDNN\cite{desplanques2020ecapa} and SincNet\cite{ravanelli2018speaker}. We randomly shuffled dataset $D_{c1}$ and divided it into 95\% for the training set and 5\% for the test set, ensuring that the two sets are non-overlapping.

\noindent\textbf{Baseline and Trigger Setup.}  We compare SPBA with the most recent speech backdoor attacks, which are as follows: (1) backdoor attack with pixel pattern (BadNets) \cite{gu2019badnets}, (2) position-independent noisy clip backdoor attack (PIBA) \cite{shi2022audio}, (3) dual adaptive backdoor attack (DABA) \cite{liu2022opportunistic}, (4) ultrasonic voice as trigger (Ultrasonic) \cite{koffas2022can}, (5) pitch boosting and sound masking (PBSM) \cite{cai2022pbsm}, and (6) voiceprint selection and voice conversion (VSVC) \cite{cai2022vsvc}. 

We proposed that the SPBA can integrate multiple triggers into a speech classifier. We established four different configurations: \textbf{(1) (w/o MGDA, K=3).} It used 3 triggers (including female, male, and angry) and optimize the neural network without MGDA. \textbf{(2) (w/o MGDA, K=5).} It used 5 triggers (including female, male, angry, sad, and happy) without MGDA. \textbf{(3) (MGDA, K=3.)} It utilized the same 3 triggers with MGDA. \textbf{(4) (MGDA, K=5.)} It incorporated the same 5 triggers with MGDA. Specifically, the utterances, including triggers, are selected from the ESD dataset\cite{zhou2022emotional}. We used the Paraformer\cite{gao2022paraformer} as the STM in Equation (2).

% 第一个表格（KWS Task）
\begin{table*}[ht]
\centering
\caption{The AV (\%), ASR (\%), and PN of baselines and SPBA on KWS task.}
\label{table: KWS}
\begin{tabular}{l|cccc}

\hline
Methods & ResNet18 & Attention-LSTM & KWS-VIT & EAT-S \\
\hline
BadNets               & 0.98/99.97/550 & 1.21/99.98/550 & 1.01/99.98/600 & 1.20/99.96/550 \\
Ultrasonic            & 2.67/97.82/350 & 2.92/97.68/400 & 3.01/96.92/400 & 2.82/97.25/400 \\
PIBA                  & 2.68/94.21/300 & 2.92/93.58/350 & 3.15/94.62/350 & 3.61/93.59/350 \\
DABA                  & 3.65/93.25/450 & 4.21/92.52/400 & 3.91/92.55/450 & 4.55/93.45/450 \\
PBSM                  & 0.58/99.98/300 & 0.54/99.88/300 & 0.72/99.94/350 & 0.66/99.87/350 \\
VSVC                  & 0.51/99.98/250 & 0.50/99.78/250 & 0.78/99.92/300 & 0.56/99.93/300 \\
\hline
\textbf{SPBA (w/o MGDA, K=3)} & 1.47/98.82/750 & 1.67/97.82/750 & 1.50/96.90/900 & 1.78/97.91/900 \\
\textbf{SPBA (w/o MGDA, K=5)} & 1.27/97.92/1000 & 1.35/97.35/1250 & 1.47/96.35/1250 & 1.19/98.35/1000 \\
\textbf{SPBA (MGDA,K=3)} & 0.42/99.92/360 & 0.53/99.15/330 & 0.74/99.65/300 & 0.69/99.75/330 \\
\textbf{SPBA (MGDA,K=5)} & 0.62/99.95/450 & 0.52/99.65/500 & 0.84/99.76/400 & 0.80/99.56/500 \\
\hline
\end{tabular}
\end{table*}

% 第二个表格（SV Task）
\begin{table*}[ht]
\centering
\caption{The AV (\%), ASR (\%), and PN of baselines and SPBA on SVs task.}
\label{table: SV}
\begin{tabular}{l|cc}
\hline
Methods & ECAPA-TDNN & SincNet \\
\hline
BadNets               & 1.04/99.85/350 & 1.26/99.80/400 \\
Ultrasonic            & 2.05/96.75/400 & 2.67/95.12/450 \\
PIBA                  & 4.16/92.15/300 & 3.95/93.01/350 \\
DABA                  & 3.98/94.05/350 & 4.65/92.81/400 \\
PBSM                  & 0.72/99.88/250 & 0.64/99.92/300 \\
VSVC                  & 0.72/99.91/250 & 0.75/99.93/300 \\
\hline
\textbf{SPBA (w/o MGDA, K=3)} & 1.44/98.01/900 & 1.21/97.21/1050 \\
\textbf{SPBA (w/o MGDA, K=5)} & 1.34/95.89/1500 & 1.14/96.77/1250 \\
\textbf{SPBA (MGDA,K=3)} & 0.84/99.55/360 & 0.77/99.25/390 \\
\textbf{SPBA (MGDA,K=5)} & 0.68/99.94/400 & 0.63/99.95/450 \\
\hline
\end{tabular}
\end{table*}

\noindent\textbf{Backdoor Training Setup.} For the KWS task, all victim models were trained using the following parameters: a batch size of 64, a training epoch of 60, and the Adam optimizer with a learning rate of 1e-4. All utterances were segmented or padded to a duration of 1 second. For the SVs task, the models were trained with the following parameters: a batch size of 64 and a training epoch of 100; the optimizer is Adam, with a learning rate that decreases from 5e-4 to 1e-4, and all utterances are segmented or padded to a duration of 3 seconds.

\noindent\textbf{Evaluation Metrics.}
The metrics include attack metrics and trigger metrics. \textit{(1) Attack metrics.} We use three metrics: Attack Success Rate (ASR), Accuracy Variance (AV), and Poisoned Number (PN) to assess the effectiveness of the backdoor attack. ASR measures the backdoor attack performance on the test dataset. AV indicates the model's prediction accuracy variance for training before and after the backdoor attacks. Compared with the same datasets, PN directly reflects the costs associated with different triggers for backdoor embedding. \textit{(2) Trigger metrics.} Trigger metrics include Mean Opinion Score (MOS) and trigger accuracy (TA), which demonstrate the effectiveness of the triggers. We use MOS to evaluate the quality of the poisoned utterances. Furthermore, we utilize the state-of-the-art open-source multimodal model, Qwen-Audio \cite{chu2023qwen}, to assess whether the emotional or timbre attributes of the poisoned samples align with the triggers. The outcome of this assessment is known as trigger accuracy (TA). The timbre and trigger prompts fed into Qwen-Audio include the poisoned samples and texts: \textit{"Given the known emotions: angry, happy, and sad, please determine the emotional category of the following speech."} and \textit{"Are the following two audio samples from the same speaker?"}. We determine the trigger samples' emotional category and timbre similarity based on the model's feedback.

\subsection{Main Results}

\noindent\textbf{Baselines Attack Results.} We present the AV, ASR, and PN values in Tables \ref{table: KWS} and \ref{table: SV}. We utilized the PN instead of the conventional metric poisoning rate (PR) to more intuitively observe the quantity of each trigger used in the backdoor attack experiments aimed at achieving the best ASR.

The tables present the backdoor attack baselines from BadNets to VSVC. The baselines utilizing perturbation triggers (including BadNets, Ultrasonic, PIBA, and DABA) exhibit AV values exceeding 1.0\% and low ASR values (below 99\% on average). This indicates that these triggers possess strong attack capabilities but lack stealthiness. Because these triggers disrupt the naturalness of speech inputs while generating poisoned samples, they significantly reduce classification accuracy during backdoor training, resulting in high AV values. Additionally, they require high PN values ranging from 300 to 500 per trigger. In contrast, methods based on element triggers cause minimal disruption to speech, leading to lower AV values. Their attack effectiveness is also superior, as shown by lower PN values ranging from 200 to 300 per trigger.

\noindent\textbf{SPBA Attack Results.} We conducted experiments with 3 and 5 triggers without using MGDA, which is equivalent to merely increasing the number of triggers. We found that the PN value for each trigger was relatively high (ranging from 250 to 350), while the ASR values were lower than the baselines. The results of the experiments conducted with 3 and 5 triggers using MGDA indicate that each trigger requires only 90 to 130 (equivalent to $450/5$ to $390/3$) poisoned samples to achieve the best ASR values under the MGDA and balanced loss. These experimental findings demonstrate that implementing the MGDA algorithm significantly enhances the attack success rate and operational efficiency of each trigger while keeping the overall poisoning rate normal.

% 修改表

\noindent\textbf{Trigger Evaluation.}
In the MOS evaluation, ten individuals were invited to participate in an auditory assessment. Each person randomly listened to 30 poisoned samples along with their corresponding clean speech samples. They were asked to judge whether the two sentences conveyed the same content, whether they sounded natural, and to provide scores ranging from 0 to 5. In the TA evaluation, we employed Qwen-Audio to calculate the accuracy of the poisoned samples and their triggers. Specifically, TA can be described using Micro-F1 scores. The final results of the evaluation are presented in Table \ref{table:MOS}. The results indicate that the poisoned samples generated by the proposed SLLM trigger demonstrate excellent speech quality and high trigger similarity.

\begin{table}[t]\footnotesize
\centering
\tabcolsep=0.15cm
\caption{The Average MOS and SER Accuracy}
\label{Table3}
\begin{minipage}{\columnwidth}
\begin{tabular}{cccccc}
\hline
\multicolumn{6}{l}{Average MOS}                                                                                                                                                             \\ \hline
Clean                                      & BadNets                    & PBSM                       & VSVC                       & SPBA                       &                            \\ \hline
4.12                                       & 3.67                       & 3.72                       & 3.94                       & 3.98                       &                            \\ \hline
\multicolumn{6}{l}{Trigger Accuracy(F1)}                                                                                                                                                    \\ \hline
\multicolumn{1}{c|}{\multirow{2}{*}{VSVC}} & \multicolumn{5}{c}{SPBA}                                                                                                                       \\ \cline{2-6} 
\multicolumn{1}{c|}{}                      & Male                       & Female                     & Angry                      & Sad                        & Happy                      \\ \hline
\multicolumn{1}{l}{0.7354}                 & \multicolumn{1}{l}{0.7498} & \multicolumn{1}{l}{0.7378} & \multicolumn{1}{l}{0.9789} & \multicolumn{1}{l}{0.9702} & \multicolumn{1}{l}{0.9688} \\ \hline
\end{tabular}
\label{table:MOS}
 \end{minipage}
\end{table}

The experimental results in Table \ref{table:MOS} indicate that our method and VSVC nearly do not compromise the quality of speech, resulting in MOS values that are close to those of the ground truth speech. In contrast, the BadNets and PBSM methods have made harmful alterations to the spectrogram and fundamental frequency of the speech, leading to a decline in speech quality. Consequently, their MOS values are lower than those of the ground truth samples. We assess emotional and speaker similarity using F1 values in the trigger accuracy. The F1 values demonstrate that the performance of the SLLM trigger aligns with the anticipated effects.

\subsection{Ablation Study}

\begin{figure}[ht]
\centering
\includegraphics[scale=0.25]{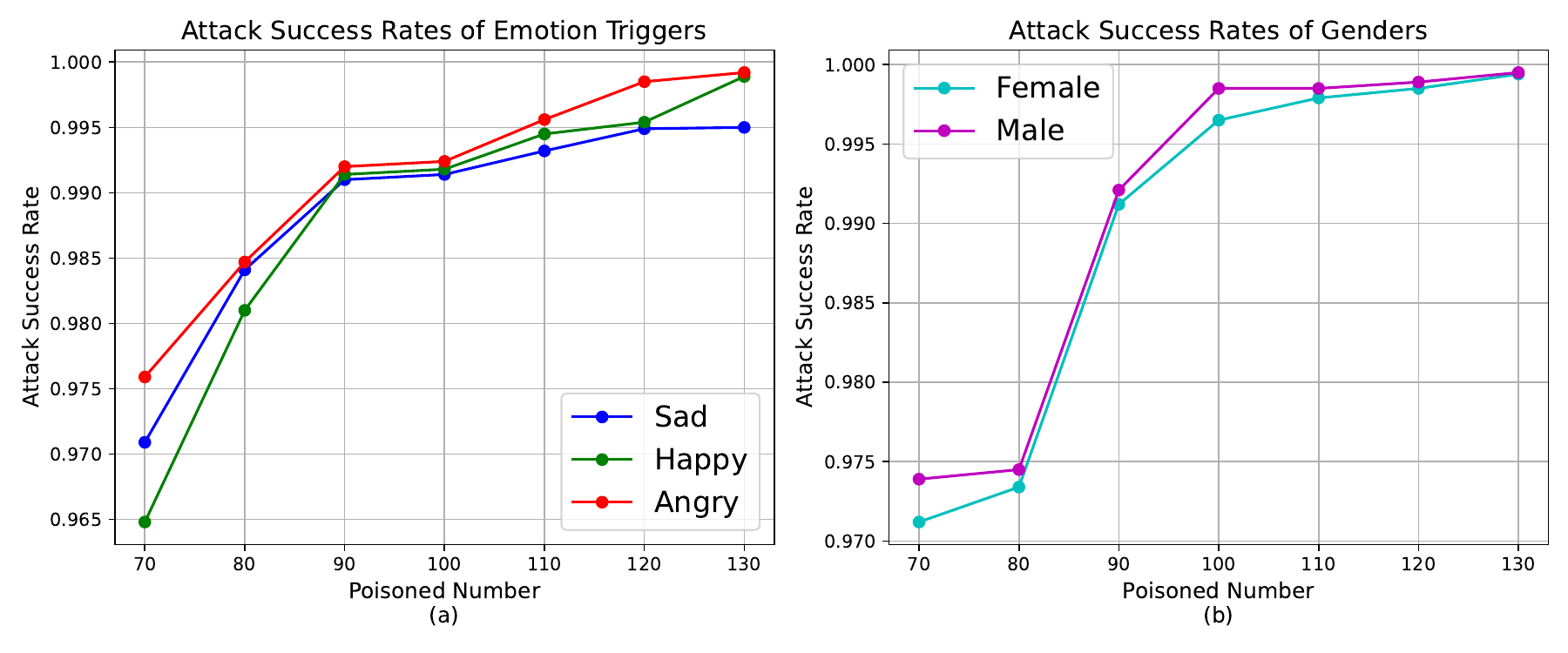}
\caption{ASR values with different emotion triggers.
}
\label{fig:emoASR}
\end{figure}

\noindent\textbf{Attack with Different Emotion Targets.} Most of the utterances in the dataset are classified as neutral speech. Therefore, we connected one of the $\{Angry, Happy, Sad\}$ as the target emotions to specific target classification labels. As shown in Figure \ref{fig:emoASR}(a), we found that intense emotions such as $\{ Angry$, and $Happy\}$ can achieve the highest ASR most quickly, while the poisoned number gradually reached 110. In other words, the classification models are more sensitive to these emotions.

\noindent\textbf{Attack with Different Gender Targets.} We used specific male and female timbres as triggers and explored the impact of these two different genders on ASR. As shown in Figure \ref{fig:emoASR}(b), in the backdoor training with multiple triggers proposed in this paper, there is no significant difference in the roles played by triggers of different genders.

\section{Conclusion}
This paper examines how embedding multiple backdoors into a DNN model simultaneously can withstand common backdoor defense strategies and proposes the SPBA method to accomplish this goal. The SPBA is a backdoor attack technique involving multiple triggers generated by the SLLM. Additionally, we use a multimodal model to assess the poisoned samples. After training with SPBA, emotional or specific gender utterances can cause the victim model to make incorrect predictions. We carried out backdoor attack experiments on two speech classification tasks. The results of these experiments highlight the remarkable effectiveness of the SPBA. Furthermore, we discovered that different emotions used as target labels lead to varying trigger efficiency. Intense emotions produce better outcomes, while triggers related to different genders play a similar role. The proposed method aims to offer insights into backdoor attacks within the speech domain.

\bibliographystyle{IEEEtran}
\bibliography{template}

% Generated by IEEEtran.bst, version: 1.14 (2015/08/26)
\begin{thebibliography}{10}
\providecommand{\url}[1]{#1}
\csname url@samestyle\endcsname
\providecommand{\newblock}{\relax}
\providecommand{\bibinfo}[2]{#2}
\providecommand{\BIBentrySTDinterwordspacing}{\spaceskip=0pt\relax}
\providecommand{\BIBentryALTinterwordstretchfactor}{4}
\providecommand{\BIBentryALTinterwordspacing}{\spaceskip=\fontdimen2\font plus
\BIBentryALTinterwordstretchfactor\fontdimen3\font minus \fontdimen4\font\relax}
\providecommand{\BIBforeignlanguage}[2]{{%
\expandafter\ifx\csname l@#1\endcsname\relax
\typeout{** WARNING: IEEEtran.bst: No hyphenation pattern has been}%
\typeout{** loaded for the language `#1'. Using the pattern for}%
\typeout{** the default language instead.}%
\else
\language=\csname l@#1\endcsname
\fi
#2}}
\providecommand{\BIBdecl}{\relax}
\BIBdecl

\bibitem{goldblum2022dataset}
M.~Goldblum, D.~Tsipras, C.~Xie, X.~Chen, A.~Schwarzschild, D.~Song, A.~M{\k{a}}dry, B.~Li, and T.~Goldstein, ``Dataset security for machine learning: Data poisoning, backdoor attacks, and defenses,'' \emph{IEEE Transactions on Pattern Analysis and Machine Intelligence}, vol.~45, no.~2, pp. 1563--1580, 2022.

\bibitem{li2022backdoor}
Y.~Li, Y.~Jiang, Z.~Li, and S.-T. Xia, ``Backdoor learning: A survey,'' \emph{IEEE Transactions on Neural Networks and Learning Systems}, 2022.

\bibitem{bagdasaryan2021blind}
E.~Bagdasaryan and V.~Shmatikov, ``Blind backdoors in deep learning models,'' in \emph{30th USENIX Security Symposium (USENIX Security 21)}, 2021, pp. 1505--1521.

\bibitem{turner2019label}
A.~Turner, D.~Tsipras, and A.~Madry, ``Label-consistent backdoor attacks,'' \emph{arXiv preprint arXiv:1912.02771}, 2019.

\bibitem{dai2019backdoor}
J.~Dai, C.~Chen, and Y.~Li, ``A backdoor attack against lstm-based text classification systems,'' \emph{IEEE Access}, vol.~7, pp. 138\,872--138\,878, 2019.

\bibitem{pan2022hidden}
X.~Pan, M.~Zhang, B.~Sheng, J.~Zhu, and M.~Yang, ``Hidden trigger backdoor attack on $\{$NLP$\}$ models via linguistic style manipulation,'' in \emph{31st USENIX Security Symposium (USENIX Security 22)}, 2022, pp. 3611--3628.

\bibitem{chen2021mitigating}
C.~Chen and J.~Dai, ``Mitigating backdoor attacks in lstm-based text classification systems by backdoor keyword identification,'' \emph{Neurocomputing}, vol. 452, pp. 253--262, 2021.

\bibitem{gu2019badnets}
T.~Gu, K.~Liu, B.~Dolan-Gavitt, and S.~Garg, ``Badnets: Evaluating backdooring attacks on deep neural networks,'' \emph{IEEE Access}, vol.~7, pp. 47\,230--47\,244, 2019.

\bibitem{liu2020reflection}
Y.~Liu, X.~Ma, J.~Bailey, and F.~Lu, ``Reflection backdoor: A natural backdoor attack on deep neural networks,'' in \emph{Computer Vision--ECCV 2020: 16th European Conference, Glasgow, UK, August 23--28, 2020, Proceedings, Part X 16}.\hskip 1em plus 0.5em minus 0.4em\relax Springer, 2020, pp. 182--199.

\bibitem{chen2017targeted}
X.~Chen, C.~Liu, B.~Li, K.~Lu, and D.~Song, ``Targeted backdoor attacks on deep learning systems using data poisoning,'' \emph{arXiv preprint arXiv:1712.05526}, 2017.

\bibitem{tran2018spectral}
B.~Tran, J.~Li, and A.~Madry, ``Spectral signatures in backdoor attacks,'' \emph{Advances in neural information processing systems}, vol.~31, 2018.

\bibitem{zhao2020clean}
S.~Zhao, X.~Ma, X.~Zheng, J.~Bailey, J.~Chen, and Y.-G. Jiang, ``Clean-label backdoor attacks on video recognition models,'' in \emph{Proceedings of the IEEE/CVF conference on computer vision and pattern recognition}, 2020, pp. 14\,443--14\,452.

\bibitem{koffas2022can}
S.~Koffas, J.~Xu, M.~Conti, and S.~Picek, ``Can you hear it? backdoor attacks via ultrasonic triggers,'' in \emph{Proceedings of the 2022 ACM workshop on wireless security and machine learning}, 2022, pp. 57--62.

\bibitem{cai2023towards}
H.~Cai, P.~Zhang, H.~Dong, Y.~Xiao, S.~Koffas, and Y.~Li, ``Towards stealthy backdoor attacks against speech recognition via elements of sound,'' \emph{arXiv preprint arXiv:2307.08208}, 2023.

\bibitem{koffas2023going}
S.~Koffas, L.~Pajola, S.~Picek, and M.~Conti, ``Going in style: Audio backdoors through stylistic transformations,'' in \emph{ICASSP 2023-2023 IEEE International Conference on Acoustics, Speech and Signal Processing (ICASSP)}.\hskip 1em plus 0.5em minus 0.4em\relax IEEE, 2023, pp. 1--5.

\bibitem{zhai2021backdoor}
T.~Zhai, Y.~Li, Z.~Zhang, B.~Wu, Y.~Jiang, and S.-T. Xia, ``Backdoor attack against speaker verification,'' in \emph{ICASSP 2021-2021 IEEE International Conference on Acoustics, Speech and Signal Processing (ICASSP)}.\hskip 1em plus 0.5em minus 0.4em\relax IEEE, 2021, pp. 2560--2564.

\bibitem{shi2022audio}
C.~Shi, T.~Zhang, Z.~Li, H.~Phan, T.~Zhao, Y.~Wang, J.~Liu, B.~Yuan, and Y.~Chen, ``Audio-domain position-independent backdoor attack via unnoticeable triggers,'' in \emph{Proceedings of the 28th Annual International Conference on Mobile Computing And Networking}, 2022, pp. 583--595.

\bibitem{liu2022backdoor}
P.~Liu, S.~Zhang, C.~Yao, W.~Ye, and X.~Li, ``Backdoor attacks against deep neural networks by personalized audio steganography,'' in \emph{2022 26th International Conference on Pattern Recognition (ICPR)}.\hskip 1em plus 0.5em minus 0.4em\relax IEEE, 2022, pp. 68--74.

\bibitem{liu2022opportunistic}
Q.~Liu, T.~Zhou, Z.~Cai, and Y.~Tang, ``Opportunistic backdoor attacks: Exploring human-imperceptible vulnerabilities on speech recognition systems,'' in \emph{Proceedings of the 30th ACM International Conference on Multimedia}, 2022, pp. 2390--2398.

\bibitem{xin2022natural}
J.~Xin, X.~Lyu, and J.~Ma, ``Natural backdoor attacks on speech recognition models,'' in \emph{International Conference on Machine Learning for Cyber Security}.\hskip 1em plus 0.5em minus 0.4em\relax Springer, 2022, pp. 597--610.

\bibitem{luo2022practical}
Y.~Luo, J.~Tai, X.~Jia, and S.~Zhang, ``Practical backdoor attack against speaker recognition system,'' in \emph{International Conference on Information Security Practice and Experience}.\hskip 1em plus 0.5em minus 0.4em\relax Springer, 2022, pp. 468--484.

\bibitem{ye2023fake}
Z.~Ye, T.~Mao, L.~Dong, and D.~Yan, ``Fake the real: Backdoor attack on deep speech classification via voice conversion,'' \emph{arXiv preprint arXiv:2306.15875}, 2023.

\bibitem{cai2022pbsm}
H.~Cai, P.~Zhang, H.~Dong, Y.~Xiao, and S.~Ji, ``Pbsm: Backdoor attack against keyword spotting based on pitch boosting and sound masking,'' \emph{arXiv preprint arXiv:2211.08697}, 2022.

\bibitem{yao2025imperceptible}
W.~Yao, J.~Yang, Y.~He, J.~Liu, and W.~Wen, ``Imperceptible rhythm backdoor attacks: Exploring rhythm transformation for embedding undetectable vulnerabilities on speech recognition,'' \emph{Neurocomputing}, vol. 614, p. 128779, 2025.

\bibitem{wang2019neural}
B.~Wang, Y.~Yao, S.~Shan, H.~Li, B.~Viswanath, H.~Zheng, and B.~Y. Zhao, ``Neural cleanse: Identifying and mitigating backdoor attacks in neural networks,'' in \emph{2019 IEEE Symposium on Security and Privacy (SP)}.\hskip 1em plus 0.5em minus 0.4em\relax IEEE, 2019, pp. 707--723.

\bibitem{wang2023neural}
C.~Wang, S.~Chen, Y.~Wu, Z.~Zhang, L.~Zhou, S.~Liu, Z.~Chen, Y.~Liu, H.~Wang, J.~Li \emph{et~al.}, ``Neural codec language models are zero-shot text to speech synthesizers,'' \emph{arXiv preprint arXiv:2301.02111}, 2023.

\bibitem{desideri2012multiple}
J.-A. D{\'e}sid{\'e}ri, ``Multiple-gradient descent algorithm (mgda) for multiobjective optimization,'' \emph{Comptes Rendus Mathematique}, vol. 350, no. 5-6, pp. 313--318, 2012.

\bibitem{simonyan2014very}
K.~Simonyan and A.~Zisserman, ``Very deep convolutional networks for large-scale image recognition,'' \emph{arXiv preprint arXiv:1409.1556}, 2014.

\bibitem{qin2017dual}
Y.~Qin, D.~Song, H.~Chen, W.~Cheng, G.~Jiang, and G.~Cottrell, ``A dual-stage attention-based recurrent neural network for time series prediction,'' \emph{arXiv preprint arXiv:1704.02971}, 2017.

\bibitem{gazneli2022end}
A.~Gazneli, G.~Zimerman, T.~Ridnik, G.~Sharir, and A.~Noy, ``End-to-end audio strikes back: Boosting augmentations towards an efficient audio classification network,'' \emph{arXiv preprint arXiv:2204.11479}, 2022.

\bibitem{wan2018generalized}
L.~Wan, Q.~Wang, A.~Papir, and I.~L. Moreno, ``Generalized end-to-end loss for speaker verification,'' in \emph{2018 IEEE International Conference on Acoustics, Speech and Signal Processing (ICASSP)}.\hskip 1em plus 0.5em minus 0.4em\relax IEEE, 2018, pp. 4879--4883.

\bibitem{desplanques2020ecapa}
B.~Desplanques, J.~Thienpondt, and K.~Demuynck, ``Ecapa-tdnn: Emphasized channel attention, propagation and aggregation in tdnn based speaker verification,'' \emph{arXiv preprint arXiv:2005.07143}, 2020.

\bibitem{cai2022vsvc}
H.~Cai, P.~Zhang, H.~Dong, Y.~Xiao, and S.~Ji, ``Vsvc: Backdoor attack against keyword spotting based on voiceprint selection and voice conversion,'' \emph{arXiv preprint arXiv:2212.10103}, 2022.

\bibitem{kim2023chatgpt}
J.~K. Kim, M.~Chua, M.~Rickard, and A.~Lorenzo, ``Chatgpt and large language model (llm) chatbots: The current state of acceptability and a proposal for guidelines on utilization in academic medicine,'' \emph{Journal of Pediatric Urology}, vol.~19, no.~5, pp. 598--604, 2023.

\bibitem{warden2018speech}
P.~Warden, ``Speech commands: A dataset for limited-vocabulary speech recognition,'' \emph{arXiv preprint arXiv:1804.03209}, 2018.

\bibitem{he2016deep}
K.~He, X.~Zhang, S.~Ren, and J.~Sun, ``Deep residual learning for image recognition,'' in \emph{Proceedings of the IEEE conference on computer vision and pattern recognition}, 2016, pp. 770--778.

\bibitem{berg2021keyword}
A.~Berg, M.~O'Connor, and M.~T. Cruz, ``Keyword transformer: A self-attention model for keyword spotting,'' \emph{arXiv preprint arXiv:2104.00769}, 2021.

\bibitem{nagrani2017voxceleb}
A.~Nagrani, J.~S. Chung, and A.~Zisserman, ``Voxceleb: a large-scale speaker identification dataset,'' \emph{arXiv preprint arXiv:1706.08612}, 2017.

\bibitem{ravanelli2018speaker}
M.~Ravanelli and Y.~Bengio, ``Speaker recognition from raw waveform with sincnet,'' in \emph{2018 IEEE spoken language technology workshop (SLT)}.\hskip 1em plus 0.5em minus 0.4em\relax IEEE, 2018, pp. 1021--1028.

\bibitem{zhou2022emotional}
K.~Zhou, B.~Sisman, R.~Liu, and H.~Li, ``Emotional voice conversion: Theory, databases and esd,'' \emph{Speech Communication}, vol. 137, pp. 1--18, 2022.

\bibitem{gao2022paraformer}
Z.~Gao, S.~Zhang, I.~McLoughlin, and Z.~Yan, ``Paraformer: Fast and accurate parallel transformer for non-autoregressive end-to-end speech recognition,'' \emph{arXiv preprint arXiv:2206.08317}, 2022.

\bibitem{chu2023qwen}
Y.~Chu, J.~Xu, X.~Zhou, Q.~Yang, S.~Zhang, Z.~Yan, C.~Zhou, and J.~Zhou, ``Qwen-audio: Advancing universal audio understanding via unified large-scale audio-language models,'' \emph{arXiv preprint arXiv:2311.07919}, 2023.

\end{thebibliography}

\end{document}